\documentclass[doublecol]{epl2} 
\usepackage[utf8]{inputenc}
\usepackage{amsmath}
\usepackage{amsfonts}
\usepackage{amssymb}
\usepackage{xcolor}
\usepackage[english]{babel}
\usepackage{fontenc}
\usepackage{graphicx}
\usepackage{MnSymbol}%

\begin{document}
\sloppy
\title{Front and Turing patterns induced by Mexican-hat-like nonlocal feedback}
\author{Julien Siebert \and Eckehard Sch\"oll}
\institute{Institut f\"ur Theoretische Physik, Technische Universit\"at Berlin, Hardenbergstra{\ss}e 36, 10623 Berlin, Germany}
\date{2014-01-20}


\abstract{
We consider the effects of a Mexican-hat-shaped nonlocal spatial coupling, i.e., symmetric long-range inhibition superimposed with short-range excitation, upon front propagation in a model of a bistable reaction-diffusion system.
We show that the velocity of front propagation can be controlled up to a certain coupling strength beyond which spatially periodic patterns, such as Turing patterns or coexistence of spatially homogeneous solutions and Turing patterns, may be induced.
This behaviour is investigated through a linear stability analysis of the spatially homogeneous steady states and numerical investigations of the full nonlinear equations in dependence upon the nonlocal coupling strength and the ratio of the excitatory and inhibitory coupling ranges.}

\maketitle
\section{Introduction}
Self-organized pattern formation in physical, chemical, or biological systems described by nonlinear reaction-diffusion equations has been widely studied \cite{HAK83,MIK94,KAP95a,KEE98}. The interplay of nonlinear chemical reactions and local diffusion can lead to a variety of spatio-temporal patterns. Stationary spatially periodic states, {\em i.e.}, Turing patterns \cite{TUR52}, 
have been observed, for instance, in the chloride-ionide-malonic acid (CIMA) reaction \cite{KEE98}.
Turing patterns can be related to morphogenesis, {\em e.g.}, the disposition for stripes in some skin patterns of fish \cite{KON10}.
Nonstationary spatio-temporal patterns include for instance propagating fronts and pulses, or
spiral waves in two-dimensional active media, like the Belousov-Zhabotinsky (BZ) chemical reaction \cite{KEE98}.

While the local diffusive coupling has been studied extensively, much less is known about the impact of nonlocal spatial interactions 
upon the dynamics of nonlinear reaction-diffusion systems. In the extreme limit of global interaction it is known that a variety of synchronization patterns, including cluster states, can be generated in oscillatory active media \cite{KUR84}.
Experimental and theoretical studies of CO oxidation on platinum surfaces \cite{FAL94,BET04a} and other catalytic processes \cite{MID92a}, in electro-chemistry \cite{PLE01} or in semiconductors \cite{MEI00b,SCH00,KEH09} have shown that global feedback can be used to control propagating waves and to generate spatially periodic patterns such as Turing patterns or travelling waves.

Recently, research has focussed on nonlocal intermediate and long-range interactions, since this has been shown to be relevant in many real-world systems ranging from chemical reactions to biology and ecology \cite{GEL10}.
Such interactions may involve a physical mechanism that couples different points in space - for example, it is often assumed that interacting cortical neurons are organized in a Mexican-hat like fashion, where short-range neighbouring cells excite each other, while cells at long-range distance inhibit each other \cite{KAN02,HUT03,ZHA14}. This nonlocal, nonmonotonic interaction can be modelled by
an integral over space weighted with a spatial kernel, which may be chosen as the superposition of two Gaussian integral kernels of opposite sign and different widths; it resembles the shape of a Mexican hat. Generally, such nonlocal couplings in the form of integrals with a spatial kernel can be derived as limiting cases of two- or three-component activator-inhibitor reaction-diffusion models with or without advection, when fast inhibitor variables are eliminated adiabatically \cite{PIS06,PET94,NIC02,TAN03,SHI04,SIE14}. In particular, asymmetric spatial kernels \cite{SIE14} arise from differential advection of some chemical species, e.g., in heterogeneous catalysis, marine biology, or ecology \cite{MID92a,ROV93}. 
It has been observed that nonlocal coupling can induce a variety of spatio-temporal patterns \cite{BEL81,SHE97a,NIC02,GEL14,COL14,LOE14,BAC14}. Such effects have been discussed in electro-chemistry \cite{MAZ97}, in the Belousov-Zhabotinsky reaction \cite{HIL01,NIC06} and magnetic fluids \cite{FRI03}, and may induce, for example, fingering \cite{PET94}, remote wave triggering \cite{CHR99}, Turing structures \cite{LI01a}, wave \cite{NIC06} instabilities or spatiotemporal chaos \cite{VAR05} in chemical systems. 
Nonlocal coupling plays also an essential rule in the formation of chimera states in discrete arrays of oscillators \cite{KUR02a,ABR04,HAG12,TIN12,MAR13,OME13,ZAK14,PAN14}.

In neuronal systems, described by the FitzHugh-Nagumo model of excitable dynamics, it was shown that traveling pulses in one spatial dimension can be suppressed by spatially nonlocal or time-delayed coupling \cite{DAH08,SCH09c}, and acceleration, deceleration, and suppression of propagating pulses as well as generation of Turing patterns, and multiple pulses and blinking traveling waves were found, depending upon the type of spatial kernel, the coupling strength, range, and scheme of the nonlocal coupling \cite{BAC14}.
The interplay between inhibitory and excitatory nonlocal interactions is particularly relevant for neurosciences, especially since propagating excitation patterns are related to the phenomenon of spreading depression that occurs during migraine or stroke \cite{DAH09a}.
In this letter, we consider the effects of a nonlocal Mexican-hat spatial feedback, which consists of a symmetric long-range inhibition superimposed with short-range excitation, upon front propagation in a generic model of a bistable reaction-diffusion system.
We show by a linear stability analysis of the spatially homogeneous steady states and by numerical integration of the full nonlinear system that the Mexican-hat spatial coupling strongly affects the propagating fronts and can accelerate, decelerate, or suppress fronts, and induce pure or mixed Turing patterns.

\section{Model}\label{sect.model}
\begin{figure}[!htbp]
  \centering
  \includegraphics[width=0.45\textwidth]{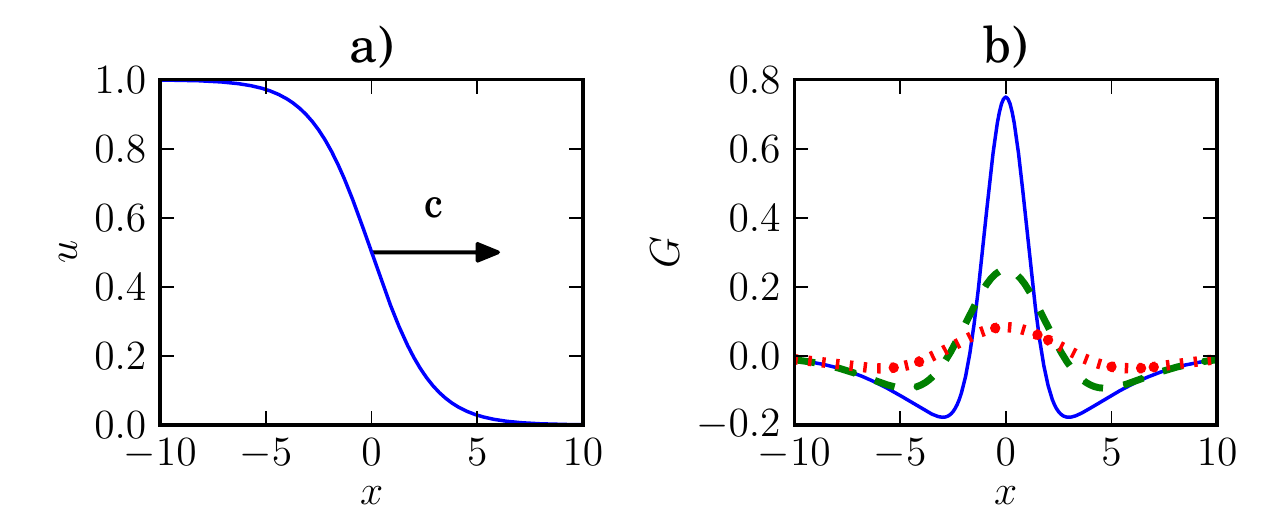}
  \caption{(a) Front shape $u(x)$ for $0 \le \alpha < 0.5$. The front velocity $c>0$ is indicated by an arrow. (b) Mexican-hat kernel $G(x)$ (Eq.~\ref{Eq.mh}). Parameters: $\rho_i=4$ and $\rho_e=1$ (solid line), $\rho_e=2$ (dashed line), $\rho_e=3$ (dotted line).}
  \label{fig.init.G}
\end{figure}

We consider a one-variable generic bistable reaction-diffusion model with a cubic nonlinearity, i.e. 
the Schl\"ogl model \cite{SCH72,SCH83a}, in one spatial dimension $x\in\mathbb R$: 
\begin{equation}
  \partial_t u = F(u) + \partial^2_x u \label{Eq.schlogl}
\end{equation}
where $u(x,t)$ is the dynamic variable, the diffusion coefficient is set equal to unity, and $F(u)$ is given by  
\begin{equation}
  F(u)=-u(u-\alpha)(u-1) \label{Eq.schlogl.F}
\end{equation}
with $0 \le\alpha \le 1$.
Eq.(\ref{Eq.schlogl}) possesses two stable spatially homogeneous solutions: $u^*=0$ and $u^*=1$ and an unstable homogeneous solution: $u^*=\alpha$.
With boundary conditions $u(-\infty,t)=1$ and $u(+\infty,t)=0$ it allows for a travelling front solution with a monotonic profile: 
\begin{equation}
  u(x,t) = \dfrac{1}{2}\left[ 1 - tanh\left( \dfrac{x-ct}{2\sqrt{2}}\right) 
\right]. \label{Eq.schlogl.front}
\end{equation}
The propagation velocity $c$ of this front is given by
$c=(1-2 \alpha)/\sqrt{2}$,
with the convention that $c>0$ if the fronts propagates in positive $x$ direction.
Fig.\ref{fig.init.G} (a) shows the front shape and direction for positive $c$.

For $0 \le \alpha<0.5$ ($c>0$), the front propagates from the domain of the globally stable state $u^*=1$ into the domain of the metastable state $u^*=0$. For $0.5 < \alpha \le 1$ ($c<0$) the front propagates from the globally stable state $u^*=0$ into the metastable state $u^*=1$ (note that the system is invariant with respect to the transformation $\alpha \to 1-\alpha, u \to 1-u$).

\section{Nonlocal feedback}\label{sect.nl}

We extend the reaction-diffusion system Eq.(\ref{Eq.schlogl}) by adding a distributed nonlocal feedback term:
\begin{equation}
  \partial_t u =  F(u)  + \partial^2_x u + \sigma \int_{-\infty}^{+\infty} G(x')u(x-x')dx', \label{Eq.nl}
\end{equation}
where $G(x)$ is the Mexican-hat nonlocal kernel:
\begin{equation}
  G(x) = \dfrac{1}{\sqrt{2\pi}}\left( \dfrac{1}{\rho_e}e^{-\tfrac{x^2}{2\rho_e^2}} - \dfrac{1}{\rho_i}e^{-\tfrac{x^2}{2\rho_i^2}}\right),
  \label{Eq.mh}
\end{equation}
with parameters $0 < \rho_e < \rho_i$. $G(x)$ is the superposition of two normalized Gaussian distributions, representing nonlocal excitatory interactions (with variance $\rho_e^2$) and nonlocal inhibitory interactions (with variance $\rho_i^2$), respectively.
Fig.~\ref{fig.init.G} (b) depicts the shape of $G(x)$ for different values of $\rho_e$ and $\rho_i$. $G(x)$ has a maximum at $x=0$ and two zeros at $x_{\pm} = \pm\rho_e\rho_i[ 2\ln{(\rho_i/\rho_e)}/ (\rho_i^2 - \rho_e^2 ) ]^{\frac{1}{2}}$.

Note that the three spatially homogeneous steady states $u^*=0$, $u^*=\alpha$, and $u^*=1$ are still solutions of Eq.(\ref{Eq.nl})
since $\int_{-\infty}^{+\infty} G(x')dx'=0$.
Their stability depends now upon the coupling strength $\sigma$. As a first step to get insight into the spatio-temporal patterns induced by the nonlocal spatial feedback, we will study the instabilities of the homogeneous steady states in dependence upon the  feedback parameters $\sigma, \rho_e, \rho_i$.

\section{Stability analysis of the homogeneous steady states}\label{sect.listan}

Performing a linear stability analysis around the homogeneous steady states ($u^*=0$, $u^*=\alpha$, $u^*=1$) by setting $u(x,t) = u^* + \delta u$ with small $\delta u = e^{-ikx}e^{\lambda t}$ we obtain the following dispersion relation from Eq.(\ref{Eq.nl}):
\begin{equation}
\lambda = F'(u^*) - k^2 + \sigma\sqrt{2\pi}\widehat{G}[k], 
\label{Eq.nl.lambda}
\end{equation}
where $\widehat{G}[k]$ is the Fourier transform of the kernel $G(x)$:
\begin{equation}
\widehat{G}[k] = \dfrac{1}{\sqrt{2\pi}}\left( e^{ \tfrac{-\rho_e^2 k^2}{2} } - e^{ \tfrac{-\rho_i^2 k^2}{2} } \right)
\label{Eq.ft.mh}
\end{equation}

\begin{figure}[!htbp]
  \centering
  \includegraphics[width=0.45\textwidth]{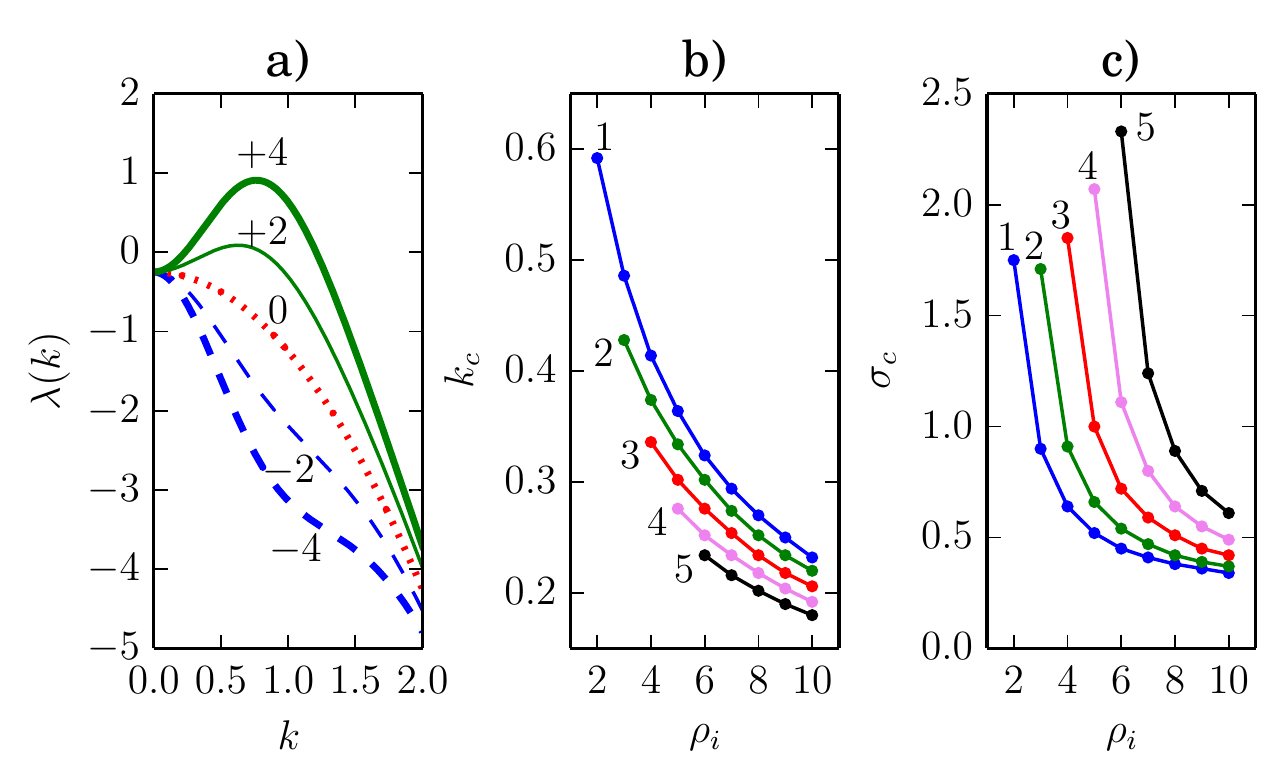}
  \caption{
    (a) Dispersion relation $\lambda(k)$ for $\rho_e=1,\rho_i=2$ and $\sigma > 0$ (solid lines); $\sigma < 0$; (dashed lines); $\sigma = 0$ (dotted line).
    (b) Dependence of the critical value $k_c$ upon $\rho_e$ (line labels) and $\rho_i$ (x-axis) for $\sigma=\sigma_c$.
    (c) Dependence of the critical value $\sigma_c$ of the Turing instability upon $\rho_e$ (line label) and $\rho_i$ (x-axis)
    Other parameters: $u^*=0$, $\alpha=0.25$
  }
  \label{fig.lambdak.sigmac}
\end{figure}

If $\lambda(k) \geq 0$, then the homogeneous steady state becomes unstable.
Fig.~\ref{fig.lambdak.sigmac} (a), shows the shape of the dispersion relation $\lambda(k)$ for different coupling strength values $\sigma$.

This stability analysis indicates that there exists a critical coupling strength $\sigma_c$ above which both stable spatially homogeneous steady states $u^*=0$ and $u^*=1$ will exhibit a Turing instability (i.e., $\lambda(k) \in \mathbb{R}$, $\lambda(k) \geq 0$ for $k\neq0$). The onset of the Turing instability is marked by the critical value $k_c$ where $\lambda(k_c) = 0$ and $\partial_k \lambda(k_c)=0$, which determines the wavelength $\Lambda \approx 2 \pi/k_c$ of the Turing pattern.
Figures~\ref{fig.lambdak.sigmac}(b) and (c) show how $k_c$  and $\sigma_c$ depend upon the ranges of the excitatory and inhibitory nonlocal feedback $\rho_e$ and $\rho_i$. Fig. \ref{fig.lambdak.sigmac}(b) agrees well with the approximate analytical relation
$k_c \propto 1/\rho_i$, i.e., the Turing wavelength is of the order of the inhibitory coupling range $\rho_i$ and widely independent of $\sigma$.
Note that the initially unstable solution $u^*=\alpha$ remains unstable for all $\sigma$.

\begin{figure}[!htbp]
  \centering
  \includegraphics[width=0.45\textwidth]{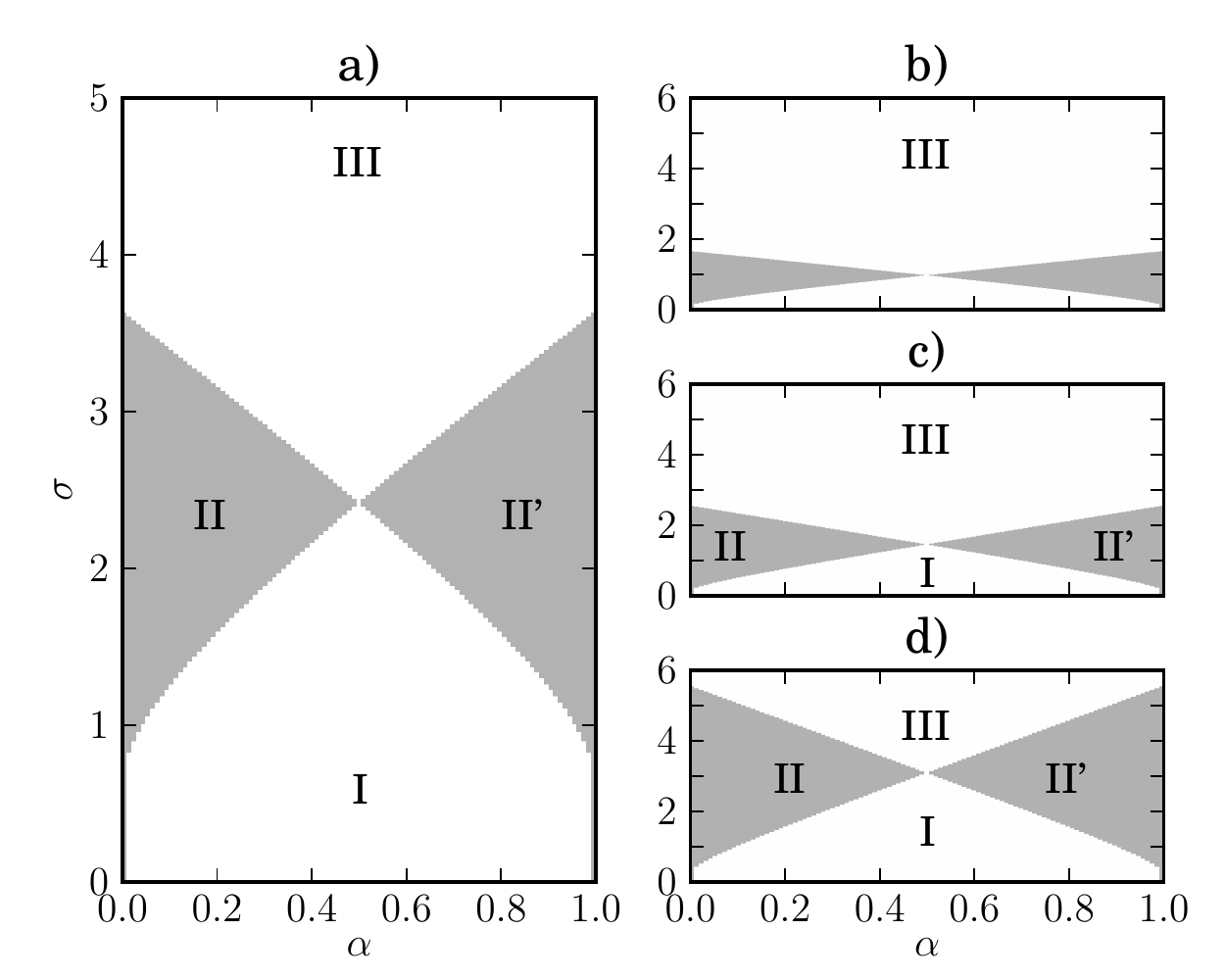}
  \caption{ Stability of the spatially homogeneous steady states $u^*=0$ and $u^*=1$ in the ($\alpha$,$\sigma$) parameter plane for
(a) $\rho_e=1,\rho_i=2$; (b) $\rho_e=1,\rho_i=4$; (c) $\rho_e=2,\rho_i=4$; (d) $\rho_e=3,\rho_i=4$. (Numerically computed from Eq.~\ref{Eq.nl.lambda}). Regime I: both solutions are stable; regime II (II'): $u^*=1$ ($u^*=0$) is stable, while $u^*=0$ ($u^*=1$) exhibits a Turing instability  (shaded light grey). Region III: both solutions exhibit a Turing instability.
  }
  \label{fig.stab}
\end{figure}

Fig.~\ref{fig.stab} shows the regimes of stability of $u^*=0$ and $u^*=1$ in the parameter space $(\alpha,\sigma)$. Regime I corresponds to the stability of both states $u^*=0$ and $u^*=1$. Regime II (II') corresponds to stability of one state and instability of the other. Regime III corresponds to instability of both states $u^*=0$ and $u^*=1$. Note that, as expected from Fig.~\ref{fig.lambdak.sigmac}(b), the locations of regimes II and III are shifted towards greater values of $\sigma$ if $\rho_e/\rho_i$ increases (Fig.~\ref{fig.stab}(b),(c),(d)). 
The Turing instability is approximately given by straight lines $\sigma(\alpha)=(k_c^2-F')/(\sqrt{2\pi}\widehat{G}[k_c])$ with 
$F'(0)=-\alpha$ and $F'(1) = \alpha - 1$ following from Eq.(\ref{Eq.nl.lambda}).

\section{Front propagation}
The linear stability analysis of the homogeneous steady states can be used to estimate the parameter ranges in which the system may exhibit spatio-temporal patterns. However, numerical integration of the full nonlinear system is necessary in order to obtain the
detailed dynamics.
Numerical integration of Eq.~(\ref{Eq.nl}), with Neumann or periodic boundary conditions, using the front profile of Eq.(\ref{Eq.schlogl.front}) as initial condition, has been performed for different sets of parameters $\alpha$, $\sigma$, $\rho_e$, and $\rho_i$.

\begin{figure}[!htbp]
  \centering
  \includegraphics[width=0.45\textwidth]{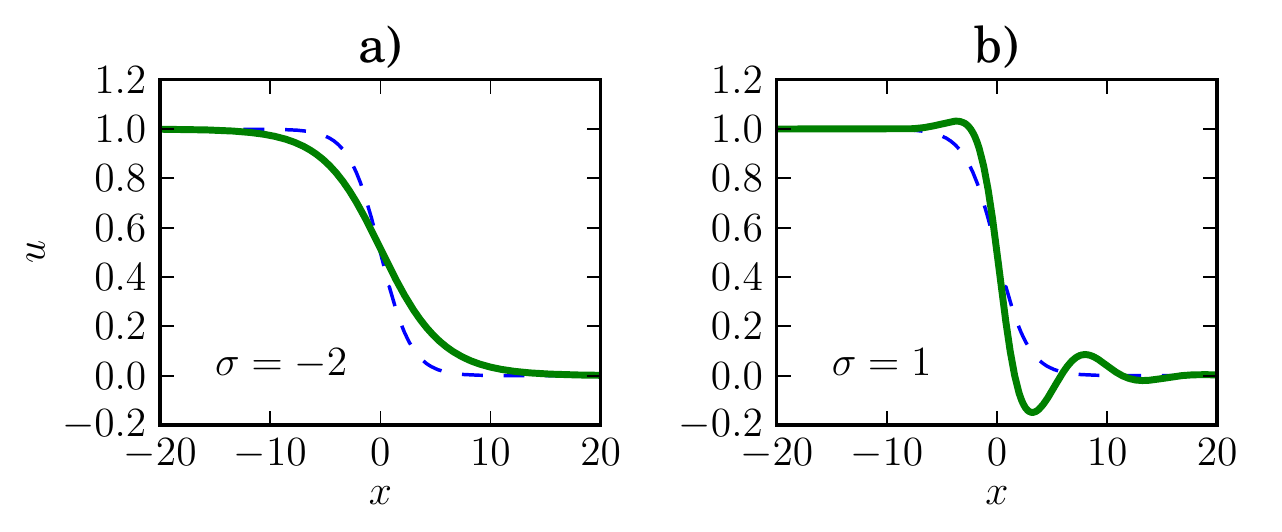}
  \caption{Front profiles $u(x)$ with (solid line) and without (dashed line) feedback. (a) Negative coupling strength ($\sigma=-2$) accelerates the front and widens its profile. (b) Positive coupling strength $(\sigma=1)$ decelerates the front and creates oscillatory tails in the profile. Parameters: $\alpha=0$, $\rho_e=1$, $\rho_i=2$. }
  \label{fig.acceldecel}
\end{figure}

We know from the linear stability analysis of the homogeneous steady states that for a coupling strength $\sigma < 0$, {\em i.e}, for short-range inhibition and long-range excitation, the system possesses two stable spatially homogeneous solutions $u^*=0$ and $u^*=1$. 
In this regime bistability and front propagation which transforms the system globally into one of those homogeneous steady states is expected.
The integration confirms this prediction, and shows that the feedback accelerates the front propagation and widens the stationary front profile (see Fig.\ref{fig.acceldecel} (a)).

If $\sigma > 0$, {\em i.e}, for long-range inhibition and short-range excitation, and if both spatially homogeneous solutions are still stable, \emph{i.e.}, in the parameter regime I of Fig.\ref{fig.stab}, the feedback decelerates the front propagation and changes the front shape by creating oscillatory tails (Fig.\ref{fig.acceldecel}. (b)) Similar effects were found for Gaussian kernels in \cite{GEL10}.

\begin{figure}[!htbp]
  \centering
  \includegraphics[width=0.45\textwidth]{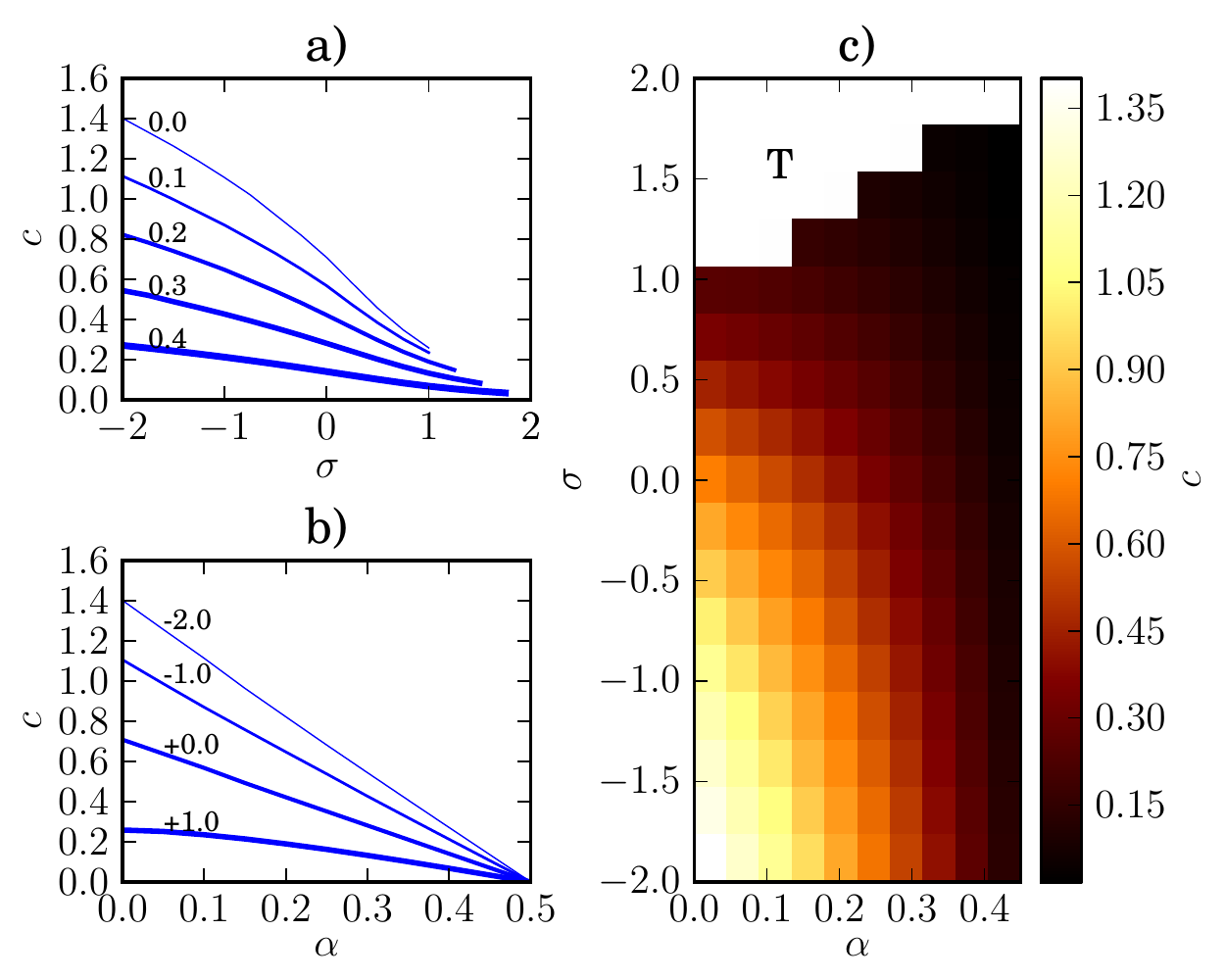}
  \caption{ Velocity $c$ of the propagating front (numerical integration).
(a) $c(\sigma)$ for different values of $\alpha$ (line labels).
(b) $c(\alpha)$ for different values of $\sigma$ (line labels).
(c) propagation velocity $c$ in the $(\alpha,\sigma)$ plane (colour-coded). $T$ denotes the emergence of Turing patterns.
Parameters $\rho_e=1,\rho_i=2$.
  }
  \label{fig.velo}
\end{figure}

The dependence of the front velocity $c$ upon of the coupling strength $\sigma$ and the parameter $\alpha$ is displayed in Fig.~\ref{fig.velo}. For a fixed value of $\alpha$, $c$ decreases until the emergence of Turing patterns (Fig.\ref{fig.velo} (a)). For a fixed value of $\sigma$, $c$ decreases with $\alpha$ (Fig.\ref{fig.velo}(b)). 
By a similar argument as in \cite{SIE14} $c$ can be calculated approximately as
\begin{eqnarray}
 c=\dfrac{ \int_0^1 F(u)\mathrm{d}u}{ \int_{-\infty}^{+\infty} (u')^2\mathrm{d}x}. \label{Eq.c.alpha<0.5}
\end{eqnarray}
Since $\int_{-\infty}^{+\infty} (u')^2  \mathrm{d}x >0$, and $\int_0^1 F(u)\mathrm{d}u = (1-2\alpha)/12>0$ for $\alpha<0.5$, the propagation velocity $c$ is positive and remains so with nonlocal coupling. However, the slope $|u'|$ of the profile is decreased for $\sigma<0$ and increased for $\sigma>0$ (Fig.~\ref{fig.acceldecel}(a) and (b), respectively). Hence $c$ increases for $\sigma<0$, while it decreases for $\sigma>0$, and the approximate linear dependence upon $\alpha$ is also explained.

\begin{figure}[!htbp]
  \centering
  \includegraphics[width=0.45\textwidth]{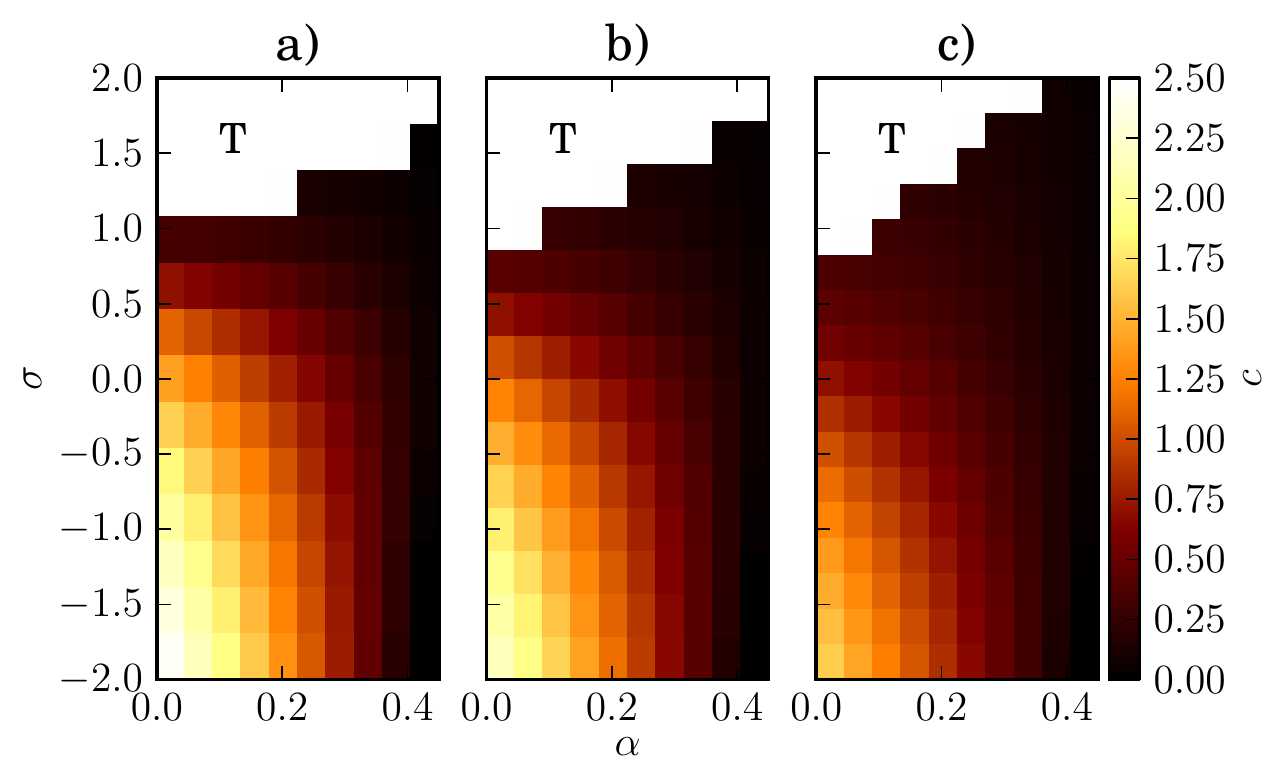}
  \caption{ Velocity $c$ of the propagating front (numerical integration) in the $(\alpha,\sigma)$ plane (colour-coded) for
$\rho_i=4$ and (a) $\rho_e=1$, (b) $\rho_e=2$, (c) $\rho_e=3$. $T$ denotes the emergence of Turing patterns.
  }
  \label{fig.velo.14_24_34}
\end{figure}

The effect of the feedback parameters $(\rho_e,\rho_i)$ characterizing the nonlocal kernel is displayed in Fig.~\ref{fig.velo.14_24_34}. Without nonlocal feedback, the front velocity depends only upon $\alpha$, see Eq.~(\ref{Eq.schlogl.velocity}), and ranges between $c=1/\sqrt{2}$ for $\alpha=0$ and $c=0$ for $\alpha=0.5$. With feedback, it depends upon the coupling strength $\sigma$ and the kernel parameters $(\rho_e,\rho_i)$. Increasing the ratio $\rho_e/\rho_i \rightarrow 1$ effectively allows for a larger coupling strength $\sigma_c$ before Turing patterns emerge. On the opposite, $\rho_e/\rho_i \rightarrow 0$ allows for much wider control over the desired front velocity, \emph{i.e.}, the range of observed front velocities $c$ becomes much broader. 

\section{Formation of Turing patterns}
\begin{figure}[!htbp]
  \centering
  \includegraphics[width=0.45\textwidth]{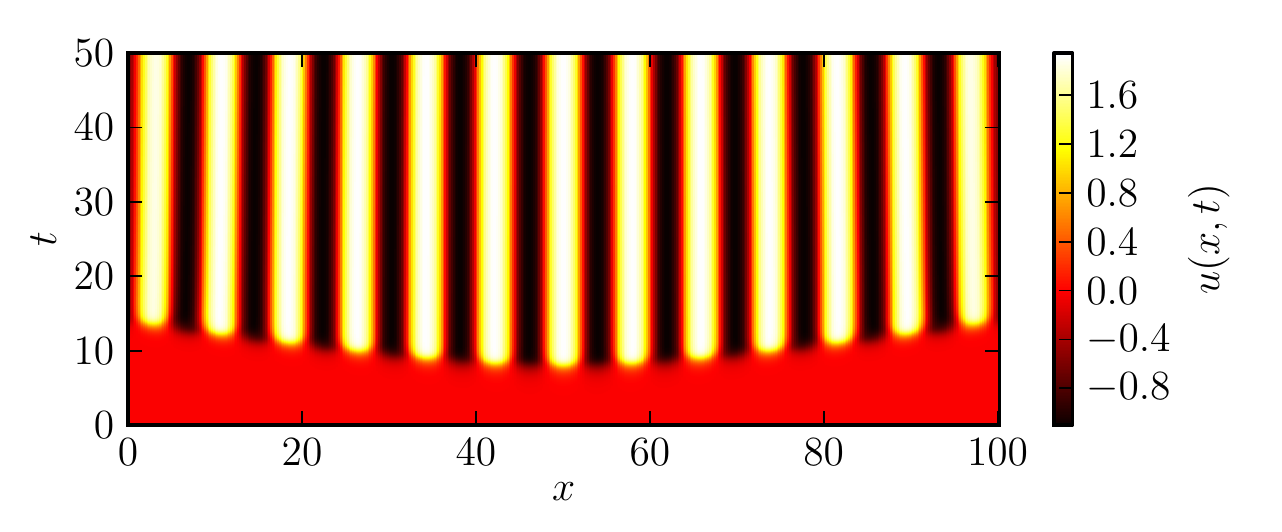}
  \caption{Emergence of Turing patterns (T) for a small perturbation of the homogeneous steady state $u_0 = 0$. The space-time plot shows $u(x,t)$ (colour-coded). Initial conditions are $u(x) = 0 \forall x \neq 50$, and $u(x=50) = 10^{-3}$. Other parameters: $\alpha = 0.25$, $\sigma = 5$, $\rho_e=1$, $\rho_i=2$.}
  \label{fig.turing.small.perturbation}
\end{figure}

Turing patterns are likely to be observed when both spatially homogeneous steady states $u^*=0$ and $u^*=1$ exhibit a Turing instability, \emph{i.e.}, for parameters in regime III of Fig.\ref{fig.stab}. Indeed, in this regime, a small perturbation of the homogeneous steady states will lead to the growth of stationary spatially periodic modes (see example in Fig.\ref{fig.turing.small.perturbation}). Note, however, that the propagating front is itself not a small perturbation. Therefore, the linear stability analysis must be complemented by numerical integration of the full equations in order to investigate the effect of nonlocal feedback. 

\begin{figure}[!hbtp]
  \centering 
  \includegraphics[width=0.45\textwidth]{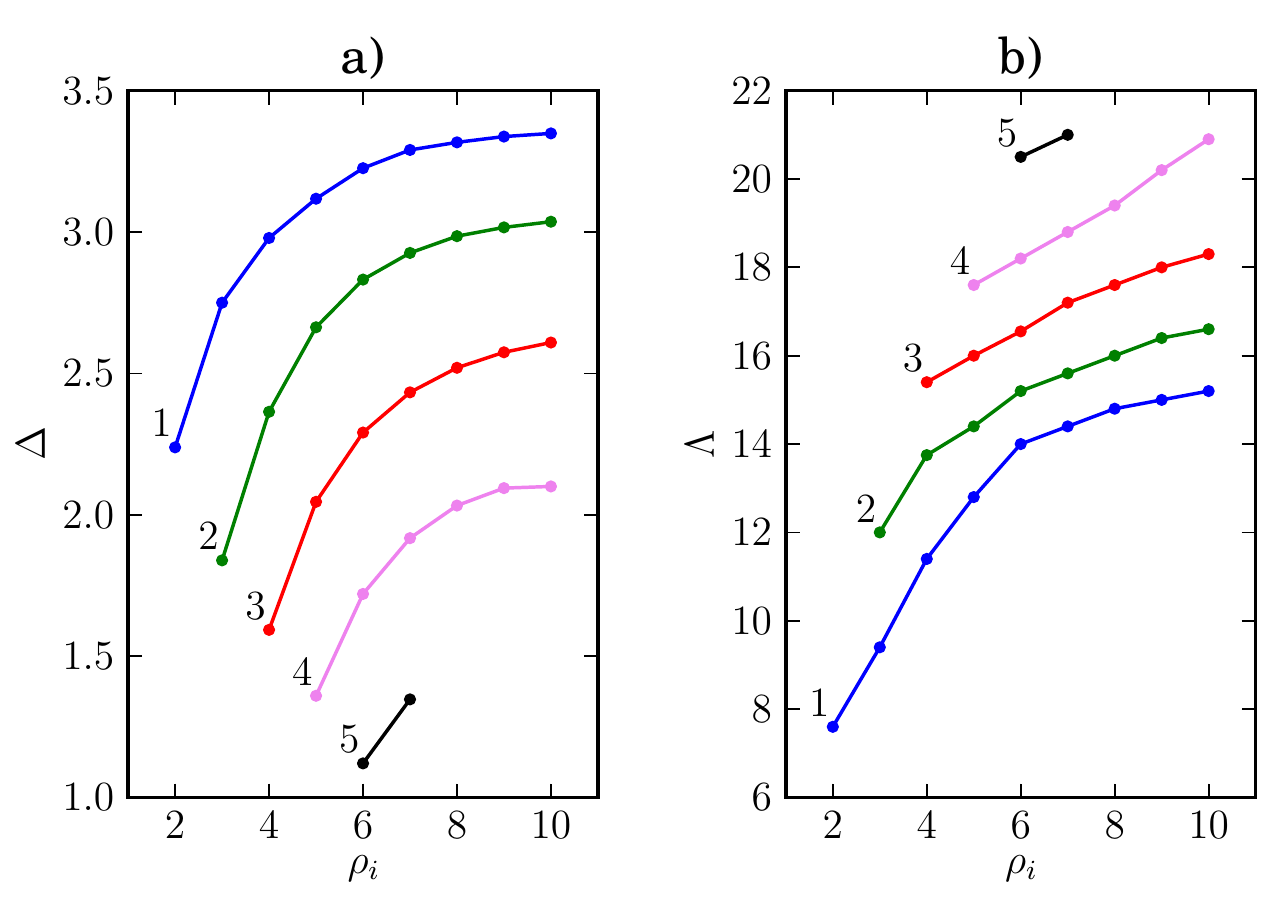}
  \caption{(a) Amplitude $\Delta$ and (b) spatial wavelength $\Lambda$ of Turing patterns for different values of $\rho_e$ (line labels) and $\rho_i$ (horizontal axis). Parameters $\sigma=10$, $\alpha=0.5$.}
  \label{fig.turing.chars}
\end{figure}

Numerical integration shows that, in the range of parameters $\alpha$, $\sigma$ close to regime III, the effect of nonlocal feedback upon front propagation is to create Turing patterns.
The spatial wavelength $\Lambda$ is approximately given by $\Lambda\approx 2\pi/k_c$ and thus depends mainly upon the kernel parameters $(\rho_e,\rho_i)$: it increases with increasing $\rho_e$ or $\rho_i$.
For fixed $(\rho_i,\rho_e)$, increasing $\sigma$ tends to significantly increase the amplitude of the patterns; the spatial wavelength is, however, not significantly affected by $\sigma$.
The effect of $\alpha$ upon the spatial wavelength and amplitude is not significant. Figure \ref{fig.turing.chars} summarizes the effect of the excitatory and inhibitory coupling radii $(\rho_e,\rho_i)$ upon the amplitude and the spatial wavelength. 

\begin{figure}[!htbp]
  \centering
  \includegraphics[width=0.45\textwidth]{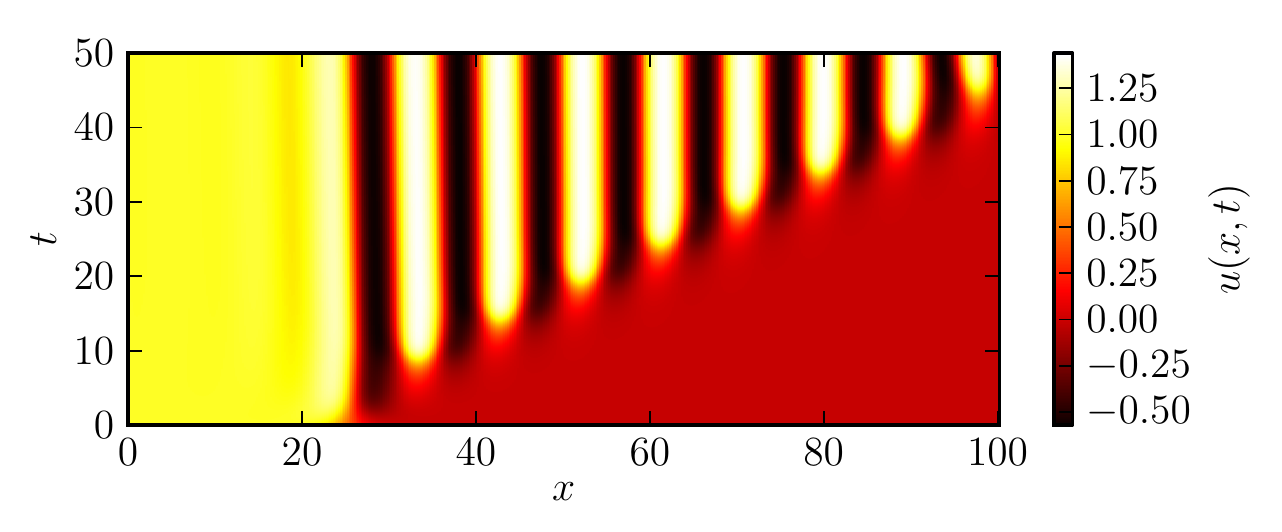}
  \caption{Spatial coexistence of Turing patterns and the homogeneous steady state $u = 1$ (T+H); $\alpha = 0.25$, $\sigma = 2.5$, $\rho_e=1, \rho_i=2$. Note that very long integration times $t > 10^4$ have been used to corroborate the stability of this pattern.}
  \label{fig.coexist}
\end{figure}

Another important phenomenon appears in the vicinity of regimes II and II', where only one homogeneous steady state is stable. In this region the nonlocal feedback leads to the spatial coexistence of Turing patterns and homogeneous steady state (Fig.\ref{fig.coexist}). 

Both pure Turing patterns and coexistence patterns are not strictly restricted to the regimes II (or II', respectively) and III.
Indeed, depending upon the system parameters, they have also been observed for parameter regime I close to the boundary with II and II'.
Figure \ref{fig.stab.stdiag} (a) summarizes the  boundaries (numerically computed) between front propagation (F), coexistence of spatially homogeneous steady state and Turing patterns (T+H) or front and Turing patterns (F+T), and pure Turing patterns (T). It usually happens that, at the boundary of regions T+H and F, the front is slightly distorted by small oscillations (Fig \ref{fig.stab.stdiag}, (e)).
Note that for $\alpha=0.5$, where the velocity of the front is zero ({\em equal areas rule} \cite{SCH00}), coexistence of both homogeneous solutions is observed for $0 \leq \sigma < \sigma_c$.

It is remarkable that the regimes I, II, and III (white and grey shading), given by linear stability analysis of the homogeneous steady states, are in reasonable agreement with the regimes F, T+H and T, respectively, of the propagating front integration (lines).

\section{Stability of spatially periodic patterns}
\begin{figure}[!htbp]
  \centering
  \includegraphics[width=0.45\textwidth]{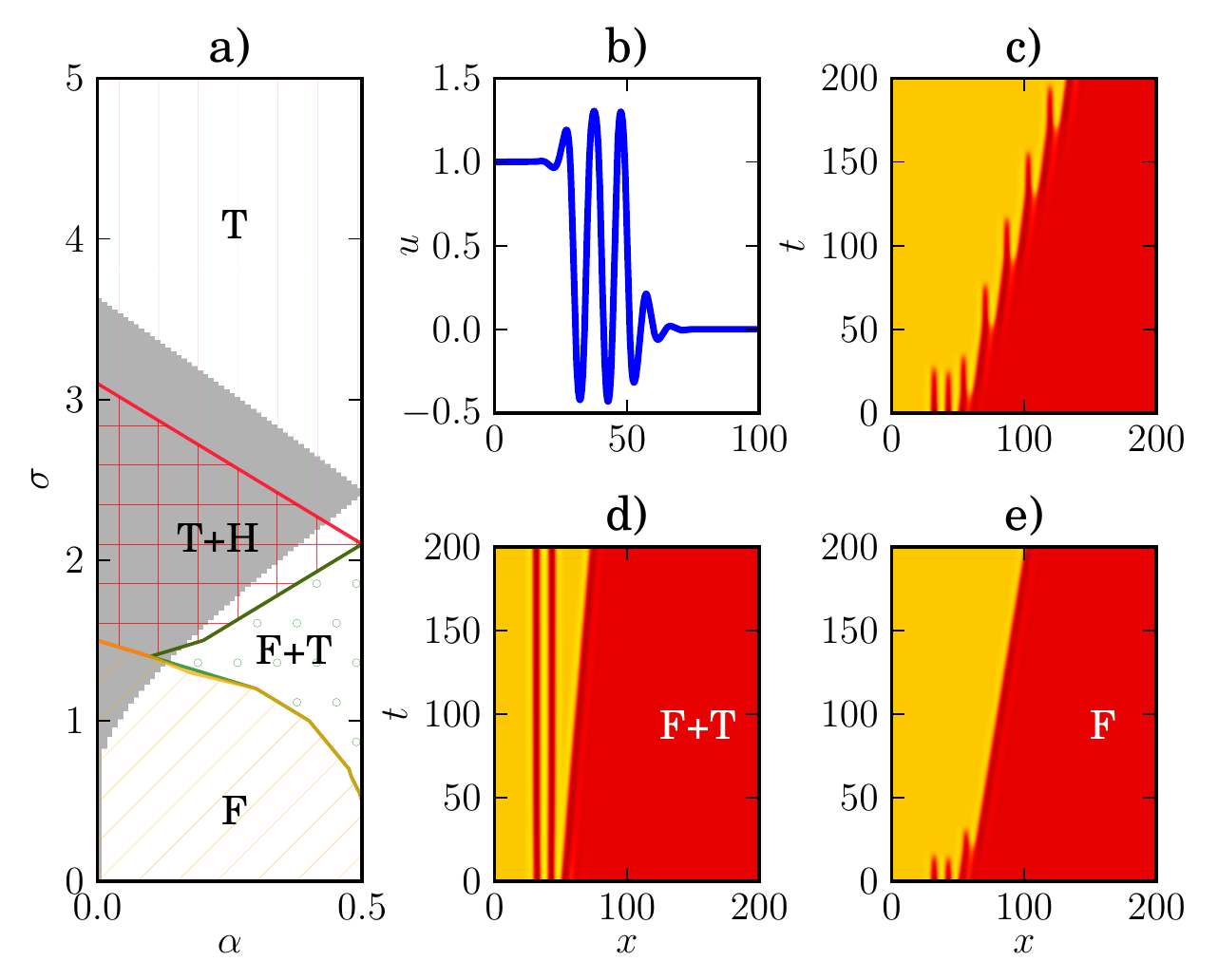}
  \caption{(a) Stability regimes in the ($\alpha,\sigma$) plane for Turing patterns (T), see Fig.\ref{fig.turing.small.perturbation}, coexistence of Turing patterns and homogeneous steady state (T+H), see Fig. \ref{fig.coexist}, propagating front (F), coexistence of propagating front and localized Turing pattern (F+T). The white and grey areas denote the stability of homogeneous steady states as in Fig.~\ref{fig.stab}(a). (b) Initial conditions used in the numerical integration. (c),(d),(e) space-time plots of $u(x,t)$ (colour-coded) for (c) $\alpha=0.1, \sigma=1.25$; (d) $\alpha=0.4, \sigma=1.25$; (e) $\alpha=0.1, \sigma=1.1$. Other parameters: $\rho_e=1,\rho_i=2$.} 
	  \label{fig.stab.stdiag}
\end{figure}

The stability of coexisting homogeneous solution and Turing patterns has been explored through numerical integration.
Starting with new initial conditions (see Fig.\ref{fig.stab.stdiag} (b)), 
the regime of front propagation (F) can be divided into a region where the initial periodic state disappears, and the front continues to propagate (Fig.\ref{fig.stab.stdiag} (e)), and into another region (called F+T) where the Turing pattern may persist without expanding (Fig.\ref{fig.stab.stdiag} (d)) and the front keeps propagating.
Figure \ref{fig.stab.stdiag} gives an overview of the different regimes.
Note that for $\alpha = 0.5$, the initial state persists in the parameter range $0 \leq \sigma < \sigma_c$.

\section{Discussions and conclusion}\label{sect.disc}
We investigate the impact of a nonlocal Mexican-hat type nonlocal spatial coupling upon a generic model of bistable media.
It is shown that long-range inhibitory and short-range excitatory feedback decreases the front propagation velocity and, for sufficient large values of the coupling strength, can provoke stationary spatially periodic states such as Turing patterns or spatial coexistence of Turing patterns and spatially homogeneous steady states.
This critical value depends upon the ratio of the spatial ranges of excitation and inhibition in the feedback kernel.
It has also been shown that stationary spatially periodic states may persist even if the coupling strength is lower than the critical value calculated from a linear stability analysis. This may be viewed as a hysteresis phenomenon associated with bistability of the homogeneous steady state and the Turing patterns.

We have systematically explored the generated patterns through a linear stability analysis of the spatially homogeneous solutions and numerical investigations of spatially inhomogeneous solutions, {\em i.e}, travelling fronts and coexistence of homogeneous solutions with spatially periodic states.
For a single propagating front, the transformation into a Turing pattern can be related to some extent to the stability of the spatially homogeneous solutions.
These results might be of relevance in neural systems, where short-range excitatory and long-range inhibitory spatial couplings are 
quite common. Our results indicate that the propagation velocity of excitations can be controlled and varied in a wider range if 
the spatial scales of short-range excitation and long-range inhibition are more separated. If these spatial scales are closer together, 
the fronts persist up to larger coupling strengths before they are suppressed by Turing patterns.
\section{Acknowledgment}
This work was supported by DFG in the framework of SFB 910 ``Control of self-organizing nonlinear systems.''

\bibliographystyle{eplbib}
\bibliography{ref}
\end{document}